\documentclass[aps,amsmath,amssymb,prb,twocolumn,showpacs,superscriptaddress]{revtex4-1}
\usepackage{color}
\usepackage{graphicx}
\usepackage{bm}
\usepackage{txfonts}
\usepackage{mathrsfs}
\usepackage{comment}
\usepackage{epsfig}
\usepackage{setspace}

\newcommand{\al}{\alpha}
\newcommand{\be}{\beta}
\newcommand{\g}{\gamma}
\newcommand{\de}{\delta}

\newcommand{\z}{\zeta}

\newcommand{\thi}{\theta}

\newcommand{\ka}{\kappa}
\newcommand{\la}{\lambda}

\newcommand{\p}{\pi}

\newcommand{\s}{\sigma}
\newcommand{\y}{\upsilon}
\newcommand{\f}{\phi}
\newcommand{\x}{\chi}
\newcommand{\w}{\omega}
\newcommand{\W}{\Omega}
\newcommand{\De}{\Delta}

\renewcommand{\y}{\psi}

\newcommand{\pd}{\partial}

\newcommand{\round}[1]{\left( #1 \right)}
\renewcommand{\square}[1]{\left[ #1 \right]}

\newcommand{\abs}[1]{\left| #1 \right|}
\newcommand{\cvec}[2]{\round{\begin{array}{c} #1 \\ #2 \end{array}}}

\newcommand{\mat}[4]{\left(\begin{array}{cc}#1&#2\\#3&#4\end{array}\right)}

\newcommand{\rvec}[2]{\round{\begin{array}{cc}#1&#2\end{array}}}

\newcommand{\ang}[1]{\left\langle #1 \right\rangle}

\newcommand{\beq}{\begin{equation}}
\newcommand{\eeq}{\end{equation}}
\newcommand{\Beq}{\begin{eqnarray}}
\newcommand{\Eeq}{\end{eqnarray}}
\newcommand{\bml}{\begin{multline}}

\newcommand{\bea}{\begin{align}}
\newcommand{\ena}{\end{align}}
\newcommand{\bsp}{\begin{split}}
\newcommand{\esp}{\end{split}}
\newcommand{\down}{\downarrow}
\newcommand{\up}{\uparrow}

\newcommand{\nn}{\nonumber}

\newcommand{\br}{{\boldsymbol r}}
\newcommand{\bM}{{\boldsymbol M}}
\newcommand{\bS}{{\boldsymbol{S}}}

\newcommand{\ez}{{\boldsymbol e_z}}
\newcommand{\bi}{{\boldsymbol i}}

\newcommand{\bJ}{{\boldsymbol J}}

\renewcommand{\bm}{{\boldsymbol m}}
\newcommand{\bk}{{\boldsymbol k}}

\newcommand{\bq}{{\boldsymbol q}}
\newcommand{\bA}{{\boldsymbol A}}
\newcommand{\bX}{{\boldsymbol X}}
\newcommand{\bY}{{\boldsymbol Y}}
\newcommand{\hO}{\hat{O}}

\newcommand{\bn}{{\boldsymbol n}}

\newcommand{\hs}{\hat{s}}

\newcommand{\vf}{\varphi}
\newcommand{\vthi}{\vartheta}
\newcommand{\bx}{\boldsymbol{x}}
\newcommand{\bt}{\boldsymbol{t}}

\newcommand{\hpsi}{\hat{\psi}}
\newcommand{\hV}{\hat{V}}
\newcommand{\ve}{\varepsilon}
\newcommand{\bs}{{\boldsymbol s}}
\newcommand{\bB}{\boldsymbol{B}}
\newcommand{\bH}{\boldsymbol{H}}
\newcommand{\bz}{\boldsymbol{z}}
\newcommand{\bb}{\boldsymbol{b}}

\newcommand{\hJ}{\hat{J}}

\newcommand{\hB}{\hat{B}}

\newcommand{\hA}{\hat{A}}

\newcommand{\tS}{\tilde{S}}

\newcommand{\sL}{\mathscr{L}}
\newcommand{\bsig}{\boldsymbol{\sigma}}
\newcommand{\bmu}{\boldsymbol{\mu}}

\begin{document}
\title{Superfluid spin transport through antiferromagnetic insulators}
\author{So Takei}
\affiliation{Department of Physics and Astronomy, University of California, Los Angeles, CA 90095, USA}
\author{Bertrand I. Halperin}
\affiliation{Department of Physics, Harvard University, Cambridge, MA 02138, USA}
\author{Amir Yacoby}
\affiliation{Department of Physics, Harvard University, Cambridge, MA 02138, USA}
\author{Yaroslav Tserkovnyak}
\affiliation{Department of Physics and Astronomy, University of California, Los Angeles, CA 90095, USA}
\date{\today}

\begin{abstract}
A theoretical proposal for realizing and detecting spin supercurrent in an isotropic antiferromagnetic insulator is reported. Superfluid spin transport is achieved by inserting the antiferromagnet between two metallic reservoirs and establishing a spin accumulation in one reservoir such that a spin bias is applied across the magnet. We consider a class of bipartite antiferromagnets with N\'eel ground states, and temperatures well below the ordering temperature, where spin transport is mediated essentially by the condensate. Landau-Lifshitz and magneto-circuit theories are used to directly relate spin current in different parts of the heterostructure to the spin-mixing conductances characterizing the antiferromagnet$|$metal interfaces and the antiferromagnet bulk damping parameters, quantities all obtainable from experiments. We study the efficiency of spin angular-momentum transfer at an antiferromagnet$|$metal interface by developing a microscopic scattering theory for the interface and extracting the spin-mixing conductance for a simple model. Within the model, a quantitative comparison between the spin-mixing conductances obtained for the antiferromagnet$|$metal and ferromagnet$|$metal interfaces is made.
\end{abstract}
\maketitle

\singlespacing

\section{Introduction}
From the early days of spin-transport electronics (or {\em spintronics}), the phenomenon of antiferromagnetism has contributed to the development of the field. It was, for instance, the discovery of antiferromagnetic interlayer coupling in a Fe/Cr/Fe system\cite{grunbergPRL86} that led shortly thereafter to the discovery of the giant magnetoresistance (GMR) effect.\cite{baibichPRL88,*binaschPRB89} Antiferromagnets (AFs) have also provided the exchange-bias effect\cite{meiklejohnPR57,*noguesJMMM99} used for enhancing magnetic stability of a neighboring ferromagnetic layer in a GMR-based spin valve. Apart from these supportive roles, antiferromagnetic materials have not yet demonstrated a prominent presence in the field of spintronics as compared to their ferromagnetic counterpart. However, the recent years have witnessed a growing interest in developing spintronic devices in which AFs play a more active role.\cite{bassetPSPIE08,*macdonalPTRSA11} Many of the key phenomena that have fueled the success of ferromagnet-based spintronics, such as GMR, current-induced torques, spin-diode effect, and magnetization switching\cite{nunezPRB06,*haneyPRB07,*urazhdinPRL07,*weiPRL07,*haneyPRL08,*xuPRL08,*gomonayPRB10,*gomonayPRB12} as well as anisotropic magnetoresistance effects,\cite{shickPRB10,*parkNATM11,*duineNATM11} have all been predicted and observed in antiferromagnetic metals. The use of AFs, in particular, has been suggested to be advantageous in reducing the critical currents for magnetization switching and achieving large magnetoresistance with relatively small applied external fields. In more recent years, theoretical works have addressed current-induced antiferromagnetic domain-wall motion\cite{wieserPRL11} and, more generally, coupled dynamics of conduction electrons with background antiferromagnetic textures.\cite{halsPRL11,*swavingPRB11,*chengPRB12,*tvetenPRL13} A very recent experiment has investigated absorption mechanisms of spin currents in antiferromagnetic metals, Ir$_{20}$Mn$_{80}$ and Fe$_{50}$Mn$_{50}$, at room temperature.\cite{merodioAPL14}

On another front, the notion of spin superfluidity in magnetic systems is currently gaining momentum.\cite{konigPRL01,*soninAP10,takeiCM13} Applying the ideas from conventional U(1) superfluidity and superconductivity to insulating magnetic systems is a focus of this endeavor. Relatively high ordering temperatures observed in magnetic insulators would entail robust manifestations of superfluidity in these systems and an exciting potential for device applications. Very recently, a large nonlocal conductance between two metallic wires mediated by a spin superfluid\cite{takeiCM13} and the manipulation of spin supercurrents using electric fields\cite{chenCM13mf} have been theoretically proposed in ferromagnetic insulators and multiferroics, respectively. In this context, antiferromagnetic insulators provide another attractive arena to study spin superfluidity using electrical controls. Within the exchange approximation, bipartite AFs host low-energy bosonic excitations with a sound-like dispersion and can support superfluid spin transport through its bulk.\cite{halperinPR69}

Ref.~\onlinecite{takeiCM13} developed a theory for nonequilibrium superfluid spin transport in ferromagnets contacted by normal metals. While planar magnetic anisotropy was necessary there for superfluid-mediated spin transport in ferromagnets, such transport is possible in both planar and isotropic AFs (with the latter subjected to a uniform magnetic field). Furthermore, in comparison to the ferromagnetic order parameter, the antiferromagnetic N{\'e}el order does not couple strongly to the magnetic field. The spin superfluid properties are thus expected to be considerably less sensitive to random magnetic fields (both external and weak stray fields) that would inadvertently break the necessary U(1) symmetry of the magnetic order. Therefore, AFs (which are relatively abundant among insulating crystals) may in practice turn out to be a more versatile and robust host for spin superfluidity.

\begin{figure}[t]
\centering
\includegraphics*[width=0.95\linewidth,clip=]{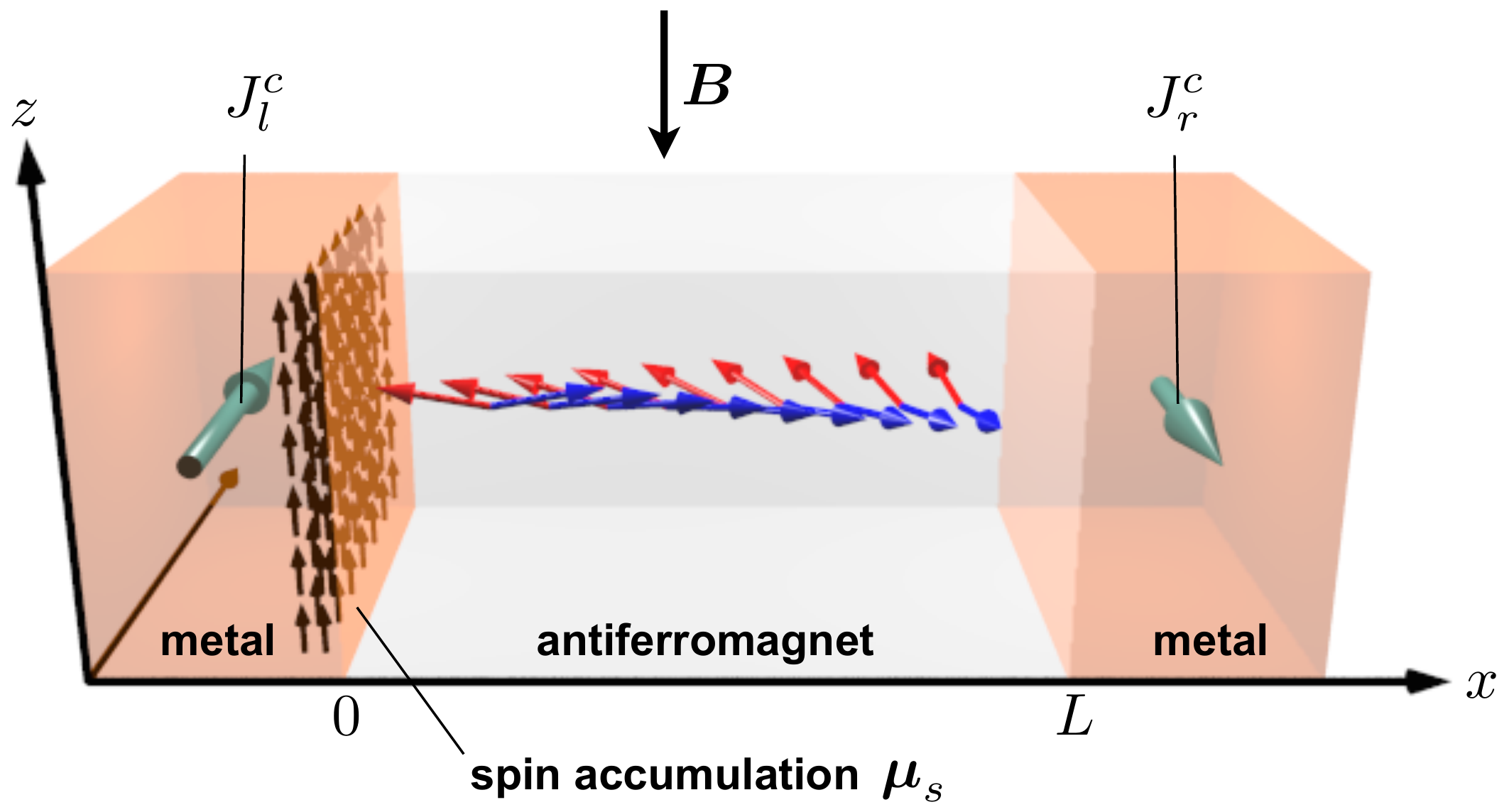}
\caption{Normal-metal$|$AF$|$normal-metal heterostructure that can be used to probe spin-superfluid transport through the AF. A charge current $J^c_l$ in the left reservoir establishes a spin accumulation $\bmu_s$ at the interface via the spin Hall effect, and the spin current pumped into the right reservoir generates a transverse charge current $J^c_r$ through the inverse spin Hall effect. Spin supercurrent through the antiferromagnetic bulk is carried by a dynamically precessing N\'eel texture. The two spin-sublattice moments are shown in the AF by approximately antiparallel arrows. Their slight canting out of the $xy$ plane arises due to a uniform magnetic field applied in the $-z$ direction and is further perturbed by the spin accumulation.}
\label{fig:setup}
\end{figure}

Specifically, in this work, we develop a theory of nonequilibrium superfluid spin transport through an isotropic antiferromagnetic insulator, and discuss its realization and detection in a spintronics device. The theory applies to a class of bipartite exchange AFs with N\'eel ground states.  We focus on the regime where thermal fluctuations of the spins are small, corresponding to the limit of temperatures well below the N\'eel ordering temperature, such that the transport is almost fully mediated by the superfluid component. We show that spin transport through an AF can be achieved and detected by sandwiching the AF with two metallic reservoirs, establishing a nonequilibrium spin accumulation (spin bias) in one of the reservoirs, and measuring spin current ejected into the other reservoir (see Fig.~\ref{fig:setup}). For an isotropic (Heisenberg) AF, we assume the presence of a weak external magnetic field collinear with the vectorial spin accumulation, such that the N\'eel order parameter is forced to lie within the plane normal to the spin-accumulation vector. In this geometry, the transmission of spin, which bears close analogy with mass superfluidity, is accomplished by the coherent rotation of the in-plane N{\'e}el order. This is contrasted with the incoherent spin transport by finite-wavelength thermal magnons, which could generally be expected to have a diffusive character in the bulk limit.

A hydrodynamic theory for the AF, in terms of the relevant slow variables parameterizing the staggered and total spin densities, is used to derive its nonequilibrium dynamics in the presence of the spin bias. We parametrize the physics of spin injection and pumping at the AF$|$metal interfaces using the spin-mixing conductance, which is obtained in close analogy with the ferromagnetic case.\cite{brataasPRL00,tserkovPRL02sp,*tserkovRMP05} Magnetic losses are taken into account by phenomenologically introducing a form of Gilbert damping that is generally applicable to the class of AFs consider in this work. Similar to spin superfluidity in ferromagnets,\cite{takeiCM13} where global spin precession and inhomogeneous magnetic textures were crucial for spin-superfluid transport, in AFs it is accomplished by the self-consistently textured and precessing N{\'e}el order. We relate the spin supercurrent flowing through the AF to the interfacial spin-mixing conductances and the Gilbert damping parameters using a combination of the bulk hydrodynamic theory and magneto-circuit theory for spin transfer at the AF$|$metal interfaces. A simple microscopic scattering theory is developed to evaluate the spin-mixing conductance at a model AF$|$metal interface, showing that it can be of the same order of magnitude as that of the ferromagnet$|$metal interface.

The paper is organized as follows. In Sec.~\ref{setup}, we review the long-wavelength theory for the bulk AF, in the absence of its coupling to the external reservoirs. Interfacial spin transfer is studied in Sec.~\ref{sfst}. A microscopic scattering theory for the spin-mixing conductance at an AF$|$metal interface is discussed in Sec.~\ref{afsmc}. We summarize the work and offer an experimental outlook in Sec.~\ref{conc}.

\section{Bulk dynamics}
\label{setup}

Let us consider insulating AFs where localized spin moments in the crystal fully compensate one another in equilibrium. In particular, we focus on bipartite AFs with two sublattices that can be transformed into each other by a symmetry transformation of the crystal. An effective long-wavelength theory for this class of AFs can be developed in terms of two slow continuum fields, $\bn(\bx)$ and $\bm(\bx)$, which parameterize the staggered (N{\'e}el) and smooth (magnetic) components of the spins, respectively, and vary slowly on the scale of the lattice spacing. The local spin directions, each belonging to one of the sublattices, can then be expressed in terms of these continuum variables as\cite{auerbachBOOK94,*sachdevBOOK99}
\beq
\label{Si}
\bS_\bi/S=\la_\bi\bn(\bx_\bi)\sqrt{1-\bm^2(\bx_\bi)}+\bm(\bx_\bi)\,,
\eeq
where $\bi$ labels the sites of the AF and $\la_\bi=\pm 1$ on the two sublattices. These continuum fields are chosen to satisfy the constraints $|\bn(\bx)|=1$ and $\bn(\bx)\cdot\bm(\bx)=0$ for all $\bx$. The presence of local N\'eel order implies $|\bm(\bx)|\ll1$. 

The AF is treated within the exchange approximation, such that the dynamic equations for $\bn$ and $\bm$ remain invariant under global spin rotations as well as space-group transformations of the crystal. This implies, in particular, their invariance under the interchange of the two sublattice spins, $\bS_A\leftrightarrow\bS_B$ ($A$ and $B$ labeling the two sublattices), such that $\bn\rightarrow-\bn$ and $\bm\rightarrow\bm$.

In order to construct the Lagrangian density $\sL_{\rm AF}$ for the AF, we follow the standard spin-coherent path-integral formulation of the problem.\cite{auerbachBOOK94} The resultant Lagrangian density can be separated into the geometric Berry-phase contribution $\sL_k$ and the dynamic contribution arising from the Hamiltonian. Inserting (\ref{Si}) into the Heisenberg Hamiltonian (for example, on a three-dimensional cubic lattice) and expanding up to quadratic order in $\bm$ and gradients of $\bn$, the Lagrangian density for the AF in the continuum form then becomes
\beq
\label{LAF}
\sL_{\rm AF}[\bm,\bn]=\sL_k-\frac{A}{2}(\pd_\mu\bn)^2-\frac{\bm^2}{2\chi}-\bb\cdot\bm\,,
\eeq
where we have introduced $\bb=\g s\bB$, in terms of the gyromagnetic ratio $\gamma$, saturated spin density $s\equiv\hbar S/\mathscr{V}$ ($\mathscr{V}$ is the volume per spin), and magnetic field $\bB$; the summation over spatial coordinates $\mu$ is implied in the second term. (The above Lagrangian expansion requires sufficiently weak fields, i.e., $b\ll\chi^{-1}$.) $A$ is the N{\'e}el-order stiffness and $\chi$ is the spin susceptibility. The kinetic (Berry-phase) term
\beq
\sL_k=s\bm\cdot(\bn\times\pd_t\bn)
\eeq
governs the canonical conjugacy between $\bn$ and $\bm$. Here, we omit a topological contribution to the action that depends on microscopic details but is irrelevant in the N\'eel phase. Integrating out the field $\bm$ in the above Lagrangian $\sL_{\rm AF}$, would reproduce the familiar Lagrangian density $\sL_N[\bn]=\chi(s\partial_t\bn+\bn\times\bb)^2/2-A(\partial_\mu\bn)^2/2$ for the isotropic N{\'e}el dynamics.\cite{andreevSPU80}

We can probe the spin-superfluid transport through the AF (of length $L$) by placing it between two metallic reservoirs as shown in Fig.~\ref{fig:setup}. A large interface in the $yz$ plane with full translational symmetry and periodic boundary conditions is assumed. The temperature is taken to be constant (and low) across the entire heterostructure, so that spin transport is driven purely by a spin bias in the absence of any thermal gradients. Each metallic reservoir is modeled as a Fermi liquid made up by spin-up and down electrons. The nonequilibrium spin accumulation, fomented, e.g. by the spin Hall effect, is introduced in the left reservoir by assigning different chemical potentials to the two spin species, $\mu_{L\up}$ and $\mu_{L\down}$, such that each species occupies the single-particle states according to the respective Fermi-Dirac distribution, $n_{L\s}(\ve)=[e^{\be(\ve-\mu_{L\s})}+1]^{-1}$. In the right reservoir, the absence of spin accumulation implies $\mu_{R\up}=\mu_{R\down}$. The spin quantization axis is taken to be parallel to the $z$ axis, and so the vectorial spin accumulation is defined as $\bmu_s=(\mu_{L\up}-\mu_{L\down})\ez\equiv\mu_s\ez$. 

\subsection{Classical dynamics for magnetic bulk}
\label{cldyn}
Let us first consider an isolated AF. Undamped Landau-Lifshitz dynamics for the N\'eel unit vector $\bn$ and the total (normalized) spin density $\bm$ can be obtained from Eq.~(\ref{LAF}) by minimizing the action subject to the nonlinear constraints $|\bn|=1$ and $\bn\cdot\bm=0$. The resultant dynamics are given by
\begin{align}
\label{ndot}
s\dot\bn&=\chi^{-1}\bm\times\bn+\bb\times\bn\,,\\
\label{mdot}
s\dot\bm&=A\bn\times\nabla^2\bn+\bb\times\bm\,.
\end{align}
The nonlinear constraints are evidently obeyed in these equations. These equations can be obtained by parametrizing the N\'eel vector with two angles $\thi$ (relative to the $xy$ plane) and $\f$ (relative to the $x$ axis), i.e., $\bn=(\cos\thi\cos\f,\cos\thi\sin\f,\sin\thi)$, and by defining two components of the total magnetization transverse to the N\'eel vector, $m_\thi$ and $m_\f$, such that $\bm=(-m_\thi\sin\thi\cos\f-m_\f\sin\f,-m_\thi\sin\thi\sin\f+m_\f\cos\f,m_\thi\cos\thi)$.

\begin{figure}[t]
\centering
\includegraphics*[width=0.5\linewidth,clip=]{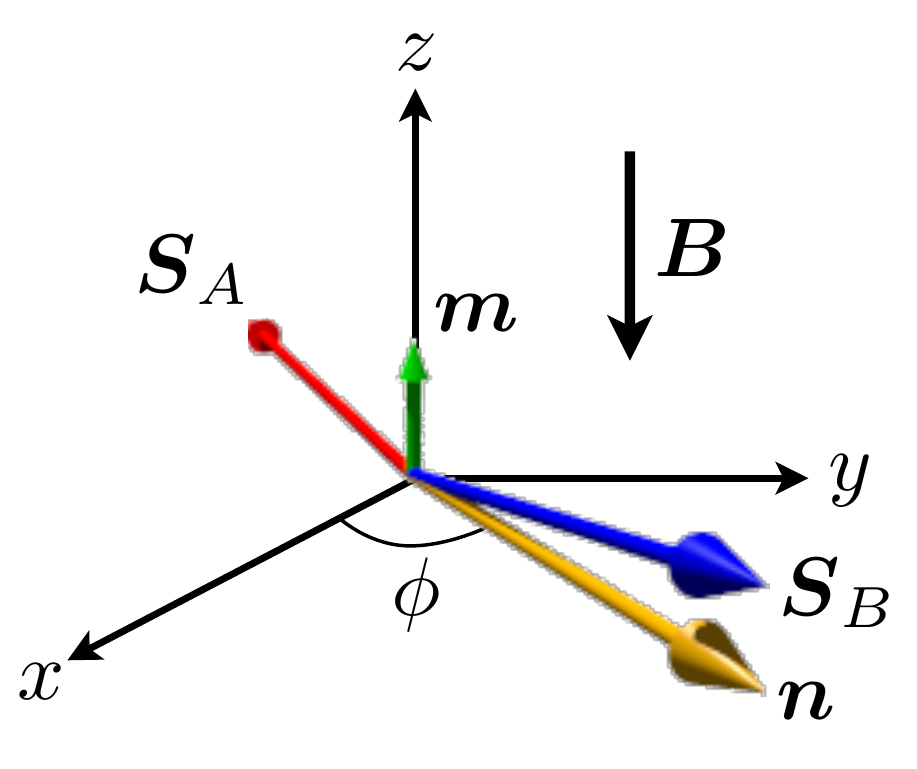}
\caption{A pictorial representation of the classical antiferromagnetic ground state in the presence of a magnetic field in the negative $z$ direction. $\bS_A$ and $\bS_B$ are the sublattice $A$ and $B$ spins, respectively. $\bn$ and $\bm$ are the corresponding N\'eel order and the net spin density (normalized by $s$). $\bn$ is taken to lie in the $xy$ plane, such that $\theta=0$.}
\label{AForder}
\end{figure}

In the presence of a uniform external field in the negative $z$ direction, i.e. $\bb=-b\ez$, the equilibrium solution to Eqs.~(\ref{ndot}) and (\ref{mdot}) is given by $\thi^{(0)}=0$, $m_{\thi}^{(0)}=\chi b$, and $m_{\f}^{(0)}=0$. The classical moments form a uniform staggered order with a slight canting of spins out of the $xy$ plane in the positive $z$ direction, which minimizes the Zeeman term in energy. The azimuthal angle $\phi$ can be arbitrary. This equilibrium state is represented pictorially in Fig.~\ref{AForder}.

\subsection{Spin waves and spin current}
\label{couplingres}

Coupling the AF to the external reservoirs perturbs this uniform static equilibrium state. In anticipation of this, we consider small deviations $\vthi$, $\xi_\thi$, and $\xi_\f$ of $\thi$, $m_\thi$, and $m_\f$, respectively, from the equilibrium solution obtained above: $m_\thi=\chi b+\xi_\thi$, $m_\f=\xi_\f$, and $\thi=\vthi$, while allowing the zero-mode coordinate $\f$ to vary smoothly over space-time. The precession of the N\'eel vector about the $z$ axis will be involved eventually in the collective (superfluid) spin transport of interest. Writing Lagrangian (\ref{LAF}) in terms of $\thi$, $\f$, $m_\thi$, and $m_\f$, and expanding up to quadratic order in $\vthi$, $\xi_\thi$, and $\xi_\f$, as well as $\dot{\phi}$ and $\boldsymbol{\nabla}\phi$, it becomes
\begin{multline}
\label{LAFexp}
\sL_{\rm AF}\approx s\left(\xi_\thi\dot\f-\xi_\f\dot\vthi\right)-\frac{A}{2}\left[(\boldsymbol{\nabla}\vthi)^2+(\boldsymbol{\nabla}\f)^2\right]\\
-\frac{\xi_\thi^2+\xi_\f^2}{2\chi}-\frac{\chi b^2}{2}\vthi^2\,.
\end{multline}
This gives the linearized Euler-Lagrange equations
\begin{align}
\label{lEL1}
s\dot\f=\chi^{-1}\xi_\thi&\,,\,\,\,s\dot \xi_\thi=A\nabla^2\f\,,\\
s\dot\vthi=-\chi^{-1}\xi_\f&\,,\,\,\,s\dot \xi_\f=-A\nabla^2\vthi+\chi b^2\vthi\,,
\label{lEL2}
\end{align}
which approximate Eqs.~(\ref{ndot}) and (\ref{mdot}). For small-amplitude fluctuations relative to a homogeneous equilibrium state, the two pairs of variables, $(\f,\xi_\thi)$ and $(\vthi,\xi_\f)$, describe two independent spin-wave branches of the AF: the former gapless with linear dispersion $\omega=cq$, in terms of the spin-wave speed $c=s^{-1}\sqrt{A/\chi}$; and the latter gapped with dispersion $\omega=\sqrt{(b/s)^2+(cq)^2}$. The direction of the applied field defines the axis of cylindrical symmetry of the system [with the gapless spin-wave branch corresponding to the associated U(1) Goldstone mode]. Therefore, in the absence of damping, the total spin angular momentum polarized along the $z$ axis is a conserved quantity. The associated spin-supercurrent density in the AF bulk can then be extracted from the continuity equation $s\dot{m}_z=-\boldsymbol{\nabla}\cdot\bJ^s$ as
\begin{equation}
\bJ^s(\bx)=-A\boldsymbol{\nabla}\f\,,
\label{Js}
\end{equation}
since $\dot{m}_z=\dot{\xi}_\theta$, in our linearized treatment. Throughout this work, we are interested only in this spin-current component, which is {\em polarized along the $z$ axis.}

\subsection{Magnetic damping}
\label{GD}

Damping of the magnetic dynamics can be phenomenologically incorporated by endowing Eqs.~(\ref{ndot}) and (\ref{mdot}) with appropriate dissipative terms. Adding viscous damping terms that are first order in time derivative, are zeroth order in spatial derivative, are time-reversal-symmetry breaking, obey a space-group symmetry flipping $\bn\to-\bn$ while $\bm\to\bm$, and satisfy the constraints $|\bn|=1$ and $\bn\cdot\bm=0$, the Landau-Lifshitz equations (\ref{ndot}) and (\ref{mdot}) are modified to
\begin{align}
\label{ndotG}
s(\dot\bn+\al\bn\times\dot\bm)&=\chi^{-1}\bm\times\bn+\bb\times\bn\,,\\
\label{mdotG}
s(\dot\bm+\al\bm\times\dot{\bm}+\al'\bn\times\dot\bn)&=A\bn\times\nabla^2\bn+\bb\times\bm\,.
\end{align}
Here, $\al$ and $\al'$ are dimensionless Gilbert-damping parameters (which can be equal or similar in simple models). Parameterizing $\bn$ and $\bm$ in terms of $\thi$, $\f$, $m_\thi$, and $m_\f$ and expanding the equations to linear order in magnetic fluctuations, the equations of motion (\ref{lEL1}) and (\ref{lEL2}) are now modified to
\begin{align}
\label{eomdamp1}
s(\dot\f-\al\dot \xi_\thi)=\chi^{-1}\xi_\thi&\,,\,\,s(\dot \xi_\thi+\al'\dot\f)=A\nabla^2\f\,,\\
s(\dot\vthi+\al\dot\xi_\f)=-\chi^{-1}\xi_\f&\,,\,\,s(\dot \xi_\f-\al'\dot\vthi)=-A\nabla^2\vthi+\chi b^2\vthi\,.
\label{eomdamp2}
\end{align}

\section{Nonlocal spin transport}
\label{sfst}
We now couple the AF to normal-metal reservoirs at its two ends. The nonequilibrium spin accumulation $\bmu_s=\mu_s\ez$ in the left reservoir causes the spins in the magnet to cant further out of the $xy$ plane (in addition to the already existing canting due to the external field). This, in turn, triggers a (right-hand) rotation of the N\'eel vector about the $z$ axis according to Eqs.~(\ref{lEL1}). This describes the injection of spin angular momentum at the left interface, which induces transport of spin through the magnet. The precessing N\'eel vector at the right interface eventually pumps spin current into the right reservoir by a reciprocal process. In the heterostructure shown in Fig.~\ref{fig:setup}, the spin accumulation in the left reservoir can be established via the spin Hall effect by driving a charge current along the $y$ axis. The spin current injected into the right reservoir can be detected through the effective electric field in the $y$ direction, which is engendered by the inverse spin Hall effect.

The magnetization dynamics in the spin-carrying AF bulk can be described by a steady-state solution to Eqs.~(\ref{eomdamp1}) and (\ref{eomdamp2}) of the form
\beq
\label{eqmansatz}
\xi_\thi=\chi s\W\,,\,\,\,\f=\vf(x)+\W t\,,\,\,\,\dot{\vthi}=\xi_\f=0\,,
\eeq
where  $\W$ is the global precession frequency of the N\'eel vector and $\vf(x)$ satisfies $A\partial_x^2\vf=\alpha's\Omega$. Both $\vf(x)$ and $\W$ must be found self-consistently from boundary conditions defined at the two AF$|$metal interfaces. 

\subsection{Interfacial spin current}

In order to establish boundary conditions for Eqs.~\eqref{eqmansatz}, we need to evaluate spin transfer across the AF$|$metal interfaces. In general, as long as the magnetic order is internally collinear, we can quantify this by a complex-valued quantity known as the (dimensionless) spin-mixing conductance (per unit area), $g^{\up\down}\equiv\Re g^{\up\down}+i\Im g^{\up\down}$. Within the magneto-circuit theory,\cite{brataasPRL00} spin transport across the interface can then be expressed in terms of the spin accumulation $\bmu_s$ in the metallic reservoir and the unit vector $\bn$ characterizing the direction of the magnetic order (which, in our case, is the unit N\'eel vector $\bn$).\cite{Note0} For the static N{\'e}el order without any magnetization, $\bm\equiv0$ , the vectorial (in spin space) spin-current density entering the AF can be written as\cite{brataasPRL00}
\beq
\label{Js0}
\bJ_0^s=\frac{\Re g^{\up\down}}{4\p}\bn\times\bmu_s\times\bn+\frac{\Im g^{\up\down}}{4\p}\bmu_s\times\bn\,.
\eeq
The subscript `0' here represents the static limit. The generalization of this result to dynamic magnetic order at lowest order in time derivatives is given by replacing
\beq
\bmu_s\rightarrow\tilde\bmu_s\equiv\bmu_s-\hbar\bn\times\dot\bn
\label{bmu}
\eeq
in Eq.~(\ref{Js0}), which can be shown directly in the scattering-matrix formalism.\cite{tserkovPRL02sp}

The static and dynamic contributions to the spin current according to Eq.~\eqref{bmu} are Onsager-reciprocal counterparts. In order to see this, let us start with the static result, Eq.~\eqref{Js0}, and invoke Onsager reciprocity to reproduce the dynamic contribution, Eq.~\eqref{bmu}. For simplicity, let us consider a homogeneous (monodomain) AF (with volume $\mathscr{V}_{\rm AF}$) in contact with a metallic reservoir (with volume $\mathscr{V}_{\rm NM}$) having a uniform spin accumulation along the $z$ axis, $\bmu_s=\mu_s\ez$. The total spin angular momentum of the magnet is $\bM=s\bm\mathscr{V}_{\rm AF}$ and that of the metallic reservoir is $\bS$. From Eq.~(\ref{LAF}), the energy of a monodomain AF is given by $F_{\rm AF}=\bM^2/2\chi s^2\mathscr{V}_{\rm AF}+\gamma\bB\cdot\bM$, which should coincide with the (mean-field) free energy at low temperatures. Although it is not essential, the following discussion is simplified if we set Gilbert damping to zero.

The dynamics for the total magnetization $\bM$, in the presence of a spin-current density $\bJ^s$ flowing into the AF, is given by
\beq
\label{Mdot}
\dot\bM=\mathscr{A}\bJ^s+\cdots\,,
\eeq
where $\mathscr{A}$ is the cross-sectional area of the interface and the ellipsis denotes terms arising due to the intrinsic Landau-Lifshitz dynamics, Eq.~\eqref{mdot}. The flow of spin current into the AF implies loss of angular momentum in the reservoir, giving the dynamics for the total spin in the reservoir of the form
\beq
\label{Sdot}
\dot\bS=-\mathscr{A}\bJ^s+\cdots\,.
\eeq
Here, the ellipsis collects terms representing the intrinsic dynamics of the reservoir spins in the absence of the coupling to the magnet (e.g. precession in external field).

Separating the spin current density $\bJ^s$ into a static term, $\bJ^s_0$, and a dynamic term, $\bJ^s_1$, induced by slow magnetization dynamics in the AF, we write $\bJ^s=\bJ^s_0+\bJ^s_1$, where $\bJ^s_0$ is given by Eq.~(\ref{Js0}). Inserting this static component into Eqs.~(\ref{Mdot}) and (\ref{Sdot}) introduces terms linear in the spin accumulation $\bmu_s$, which, by definition, is proportional to the force ${\boldsymbol f}_\bS$ conjugate to $\bS$: ${\boldsymbol f}_\bS\equiv-\de_\bS F_{\rm NM}\equiv-\bmu_s/\hbar$, where $F_{\rm NM}$ is the free energy of the normal-metal reservoir. Onsager reciprocity thus dictates an additional term in Eq.~(\ref{Sdot}) of the form:
\beq
\label{Sdot2}
\dot\bS=-\mathscr{A}\bJ^s_0+\mathscr{A}\bn\times\square{\frac{\Re g^{\up\down}}{4\p}\bn\times{\boldsymbol f}_\bM+\frac{\Im g^{\up\down}}{4\p}{\boldsymbol f}_\bM}\,,
\eeq
where ${\boldsymbol f}_\bM\equiv-\de_\bM F_{\rm AF}$ is the force conjugate to $\bM$. Noting that, according to Eq.~\eqref{ndot}, the N{\'e}el dynamics obey $\dot\bn=-{\boldsymbol f}_\bM\times\bn$, we immediately identify the full expression for the spin-current density flowing through the interface:
\beq
\label{Jsfull}
\bJ^s=\frac{\Re g^{\up\down}}{4\p}\bn\times\tilde\bmu_s\times\bn+\frac{\Im g^{\up\down}}{4\p}\tilde\bmu_s\times\bn\,.
\eeq
The term $\propto\Re g^{\up\down}$ describes the dissipative component of the interfacial spin transfer, which is analogous to Andreev reflection at superconductor$\mid$normal-metal interfaces.

\subsection{Linear response}

We are now ready to complement Eqs.~\eqref{eqmansatz} with the appropriate boundary conditions for the spin-current continuity, in the linear response to spin bias $\bmu_s$. Focusing on spin transfer in the $z$ direction (in spin space), the $\propto\Im g^{\up\down}$ term in Eq.~\eqref{Jsfull} can be disregarded, and we henceforth denote $\Re g^{\up\down}$ simply by $g^{\up\down}$. The spin current injected into the magnet at the left interface and the spin current ejected out of the magnet at the right interface are thus respectively given by
\begin{equation}\begin{aligned}
J^{s}_l&=\frac{g^{\up\down}_l}{4\p}(\mu_s-\hbar\W)=-\left.A\partial_x\phi\right|_{x=0}\,,\\
J^{s}_r&=\frac{g^{\up\down}_r}{4\p}\hbar\W=-\left.A\partial_x\phi\right|_{x=L}\,,
\label{JsLR}
\end{aligned}\end{equation}
where we used Eq.~\eqref{Js} for the collective spin-current density on the magnetic side at each interface. ($L$ is the length of the AF along the transport direction.) Here, subscripts $l$ and $r$ label the spin current at the left and right interfaces, respectively. Equation $A\partial_x^2\f=\alpha's\Omega$ is now solved together with the boundary conditions \eqref{JsLR}, in order to find the profile for phase $\f$ along the $x$ axis, global precession frequency $\Omega$, and the associated spin current throughout our structure.

In the absence of Gilbert damping in the bulk, the spin current is continuous throughout, so that $J^s_l=J^s_r$, and we find for the precession frequency and the spin-current density flowing through the magnet
\beq
\label{mainresnodamp}
\W=\frac{\mu_s}{\hbar}\frac{g^{\up\down}_l}{g^{\up\down}_l+g^{\up\down}_r}\,,\,\,\,J^s=\frac{\mu_s}{4\p}\frac{g^{\up\down}_lg^{\up\down}_r}{g^{\up\down}_l+g^{\up\down}_r}\,.
\eeq
In the presence of magnetic damping, the spin-current loss $\De J^s\equiv J_l^s-J^s_r$ along the AF satisfies
\beq
\label{DeJs}
\De J^s=-\int_0^Ldx\,\pd_x J^s(x)=\int_0^Ldx\,A\partial_x^2\phi=\al's\W L\,.
\eeq
We then obtain using (\ref{JsLR}) and (\ref{DeJs}) that
\beq
\label{mainres}
\W=\frac{\mu_s}{\hbar}\frac{g^{\up\down}_l}{g^{\up\down}_l+g^{\up\down}_r+g_\al}\,,\,\,\, J^s_r=\frac{\mu_s}{4\p}\frac{g^{\up\down}_lg^{\up\down}_r}{g^{\up\down}_l+g^{\up\down}_r+g_\al}\,,
\eeq
where $g_\al\equiv4\p\al' sL/\hbar$.

The above results, which are central to this work, are fully analogous to those obtained for the easy-plane ferromagnet.\cite{takeiCM13} Assuming the spin-mixing conductances at the two interfaces are similar, i.e. $g^{\up\down}\equiv g^{\up\down}_l\sim g^{\up\down}_r$, we define from Eqs.~(\ref{mainres}) the length scale $L_\al\equiv\hbar g^{\up\down}/4\p\al' s$. For $L\gtrsim L_\al$, the magnetic damping in the bulk is important and the spin-current ejected into the right reservoir decays inversely with the length of the AF. For $L\ll L_\al$, on the other hand, magnetic losses are negligible, and we recover Eqs.~(\ref{mainresnodamp}). 

\section{Microscopic spin transfer}
\label{afsmc}

The efficiency of the spin transfer process at an AF$|$metal interface is quantified using the spin-mixing conductance $g^{\up\down}$, which is both obtainable from microscopic theory and directly measurable in experiments. In this section, we follow the scattering-matrix formalism of Ref.~\onlinecite{brataasPRL00} to determine $g^{\up\down}$ for the AF$|$metal interfaces. Imagine an electron impinging on an antiferromagnetic insulator in a perfectly uniform N\'eel state. If an electron spin is exchange-coupled to the antiferromagnetic spins at the interface, such that the average magnetization exposed to electrons is zero, one may na{\"i}vely expect a vanishing spin transfer. We show below, however, that the spin-mixing conductance, per unit area, is generally nonvanishing in the thermodynamic limit, even for a perfectly uniform AF, due to the interference of the interfacial scattering channels. In the following, we consider a simple model that allows us to illustrate this point in detail.

\subsection{Scattering formalism}
\label{ScatTh}
\begin{figure}[t]
\centering
\includegraphics*[width=0.9\linewidth,clip=]{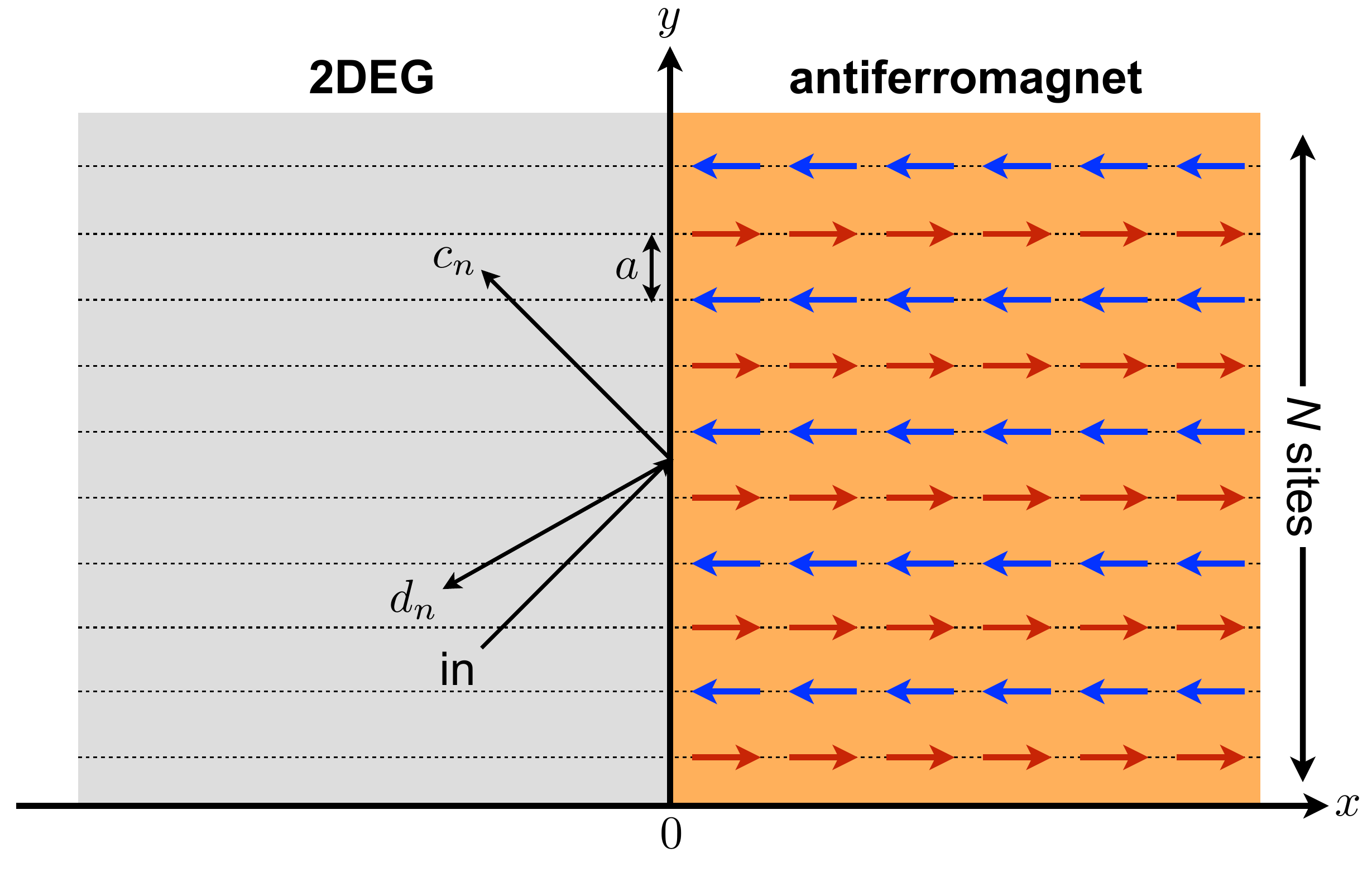}
\caption{2DEG in contact edge-on with a 2D antiferromagnetic insulator. The space is discretized in the $y$ direction as shown by dashed lines. Tight-binding dispersion is assumed in the $y$ direction with a bandwidth $2W$ (see text for details). Here, we have sketched a scattering process, in which an electron in transverse channel $n$ incoming from the left (labeled ``in") undergoes specular ($c_n$) and Umklapp ($d_n$) reflections.}
\label{mixcond}
\end{figure}

Let us consider a semi-infinite two-dimensional electron gas (2DEG) in contact edge-on with a semi-infinite 2D magnetic insulator, as shown in Fig.~\ref{mixcond}. In our model, we take $x$ as the continuum coordinate (in the transport direction), and discretize the transverse coordinate $y$ into even number of sites $N$ labeled by $i_y$. Tight-binding dispersion (with bandwidth $2W$) is assumed in the transverse direction. A large potential barrier $U_0$ is imposed for $x>0$, such that all modes entering the magnet at the Fermi level are rendered evanescent. In order to calculate $g^{\uparrow\downarrow}$, we consider a static staggered order in the AF. The specific order is modeled by a translationally-invariant arrangement of spins in the transport (i.e., $x$) direction, which are staggered along the $y$ axis, as sketched in Fig.~\ref{mixcond}. We orient both the spin-quantization axis of the electrons and the N\'eel order along the $x$ axis. The effect of the underlying magnetic order on the evanescent electrons is modeled by subjecting them to an oscillating exchange field that modulates the insulating potential: $U=U_0+\s\eta_0(-1)^{i_y}$. Here, $\s=\pm$ corresponds to up- and down-spin electrons along the N\'eel vector (pointing in the positive $x$ direction), respectively, and $\eta_0$ is the exchange-energy scale: it is taken to be constant, such that electron spins are coupled uniformly to the magnetic order. Within this simple model, we are not allowing for any spin-flip processes at the interface.

For periodic boundary condition in the $y$ direction, the transverse component of the wave function can be expanded in terms of the complete eigenbasis: $\f_n(i_y)=\sqrt{1/N}\exp[i(2\p ni_y/N)]$, where $n$ runs from $0$ to $N-1$. The scattering state for $x<0$ originating from the 2DEG can then be written as a superposition of mode $n$ impinging on the interface along with a set of reflected modes:\cite{nazarovBOOK09} 
\beq
\label{statesL}
\y^<_{nE\s}(x,i_y)=\frac{\f_{n}(i_y)e^{ik^<_n x}}{\sqrt{2\p\hbar v^<_n}}+\sum_{n'}r^\s_{n'n}\frac{\f_{n'}(i_y)e^{-ik^<_{n'} x}}{\sqrt{2\p\hbar v^<_{n'}}}.
\eeq
Here, a particular scattering state is denoted by the transverse-mode label $n$, energy $E$, electron spin $\s$, and $<$ stands for the region left of the interface. The momentum eigenvalues are given by $k^<_n=\sqrt{2m(E-E_n)}/\hbar$, the velocity by $v^<_n=\hbar k^<_n/m$, $E_n=-W\cos(2\p n/N)$ denotes the tight-binding dispersion in the transverse direction, and $m$ is the electron mass in the transport direction. The key quantity of interest here is the reflection coefficient, $r^\s_{n'n}$. The area-integrated (dimensionless) spin-mixing conductance, $G^{\up\down}\equiv\mathscr{A}g^{\up\down}$, is then given by\cite{brataasPRL00}
\beq
\label{gudformula}
G^{\up\down}=\sum_{nn'}\round{\de_{nn'}-r^\uparrow_{nn'}r^{\downarrow*}_{nn'}}\,,
\eeq
where the summation is performed over all propagating Fermi-level modes in the 2DEG.

\subsection{The Umklapp channel}

The scaling of $G^{\up\down}$ with the system size $N$ is intimately tied to the Umklapp scattering channel for electrons, which opens due to the staggered antiferromagnetic order. The $2a$ periodicity arising from the staggered antiferromagnetic order couples momentum modes that differ by wave number $\De k=\p/a$. Therefore, the reflection coefficients can in general be written as $r^\s_{n'n}=c^\s_{n}\de_{n'n}+d^\s_{n}\de_{n'\bar n}$, where $\bar n=(n+N/2)\mbox{ mod }N$. The second term represents an Umklapp channel that couples modes that are related by reciprocal lattice vectors of the magnetic Brillouin zone. The relation between reflection coefficients corresponding to the up- and down-spin electrons can be determined as follows. We first note that the scattering amplitudes for the down-spin electrons can be obtained from those of the up-spin electrons by rotating the cylinder (our spatial domain subject to the periodic boundary condition) by the lattice constant $a$. Under this rotation, the reflection coefficient of a process that couples two momenta differing by $\De k$ gains an additional multiplicative phase factor of $e^{i\De ka}$. Since the term proportional to $c^\s_{n}$ corresponds to reflection processes with $\De k=0$, we expect $c^\uparrow_{n}=c^\downarrow_{n}$. On the other hand, the coefficient for the Umklapp channel, $d^\s_{n}$, describes reflection processes where $\De k=\p/a$. We then expect a $\p$ phase shift and, therefore, $d^\uparrow_{n}=-d^\downarrow_{n}$. We thus find the spin-dependent reflection coefficients to have the general form:
\beq
r^\s_{n'n}=c_{n}\de_{n'n}+\s d_{n}\de_{n'\bar n}\,.
\eeq

In the absence of propagating modes on the right side, the spin-diagonal conductances should vanish, which, in turn, implies $|c_n|^2+|d_n|^2=1$ for all $n$ that are propagating at the Fermi level. Then, using Eq.~(\ref{gudformula}), the spin-mixing conductance becomes
\beq
G^{\up\down}=2\sum_{n}|d_n|^2\,,
\eeq
where the sum is performed over all transverse modes that have Umklapp channel available at the Fermi level. For this, we need $E>0$; if $E>W$, furthermore, all modes would participate in spin transfer. Since $|d_n|^2$ is typically nonzero for all relevant modes $n$, we see that $G^{\up\down}$, even for a purely uniform AF, would scale as $N$. We now substantiate this result using a microscopic calculation.

\subsection{Microscopic calculation}

We now return to carrying out the calculation for the set-up of Sec.~\ref{ScatTh}. The Hamiltonian on the magnetic side can be written in the $(n,n')$ basis, spanned by $\f_n(i_y)$, as
\beq
H^\s_{nn'}=\round{-\frac{\hbar^2\pd_x^2}{2m}+U_0+E_n}\de_{nn'}+\s \eta_0\de_{\bar nn'}\,.
\eeq
The off-diagonal term (proportional to $\eta_0$) couples $n$ and $\bar n$ modes, i.e. it represents the Umklapp channel. Diagonalizing in the ($n,\bar n$) subspace, the scattering state on the magnetic side can then be written as
\beq
\y^>_{nE\s}(x,i_y)=\sum_{n'}\frac{t^\s_{n'n}}{\sqrt{2\p\hbar v^>_{n'}}}\z_{n'}(i_y)e^{ik^>_{n'}x}\,,
\eeq
where $k^>_{n}=\sqrt{2m[E-U_0-s_n(E_n^2+\eta_0^2)^{1/2}]}/\hbar$ is the appropriate propagating or evanescent wave number and $v^>_{n}=\hbar k^>_{n}/m$. Here, $s_n=1$ for $0\le n< N/2$ and $s_n=-1$ for $N/2\le n< N$. The eigenfunctions are given by $\z_n(i_y)=[a_{n}\f_n(i_y)+\s b_{n}\f_{\bar n}(i_y)]e^{ik^>_nx}$ for $0\le n< N/2$ and $\z_n(i_y)=[\s b_{n}\f_n(i_y)-a_{n}\f_{\bar n}(i_y)]e^{ik^>_nx}$ for $N/2\le n< N$, where $a_{n}=\xi_{n}/(\xi^2_{n}+\eta_0^2)^{1/2}$, $b_{n}=\eta_0/(\xi^2_{n}+\eta_0^2)^{1/2}$, and $\xi_{n}=E_n+(E_n^2+\eta_0^2)^{1/2}$. 

For a given incoming channel $n$, only two reflection and two transmission amplitudes are nonzero: the momentum-conserving amplitudes $r_{nn}$ and $t_{nn}$ and the Umklapp amplitudes $r_{\bar nn}$ and $t_{\bar nn}$. Imposing the continuity of the wave function and its derivative at the interface, we obtain four equations for these amplitudes,
\begin{equation}\begin{aligned}
\label{ctymc}
1+r^\s_{nn}&=\al^+_{n\s}t^\s_{nn}+\be^+_{n\s}t^\s_{\bar nn}\,,\\
1-r^\s_{nn}&=\al^-_{n\s}t^\s_{nn}+\be^-_{n\s}t^\s_{\bar nn}\,,\\
r^\s_{\bar nn}&=\be^+_{\bar n\s}t^\s_{nn}+\al^+_{\bar n\s}t^\s_{\bar nn}\,,\\
-r^\s_{\bar nn}&=\be^-_{\bar n\s}t^\s_{nn}+\al^-_{\bar n\s}t^\s_{\bar nn}\,,
\end{aligned}\end{equation}
where 
\begin{equation}\begin{aligned}
\al^l_{n\s}&=\thi_{n}\round{\frac{k_n^<}{k^>_{n}}}^{l/2}a_{n}+\s\thi_{\bar n}\round{\frac{k_{n}^<}{k^>_{n}}}^{l/2}b_{n}\,,\\
\be^l_{n\s}&=-\thi_{n}\round{\frac{k_n^<}{k^>_{\bar n}}}^{l/2}a_{\bar n}+\s\thi_{\bar n}\round{\frac{k_{n}^<}{k^>_{\bar n}}}^{l/2}b_{\bar n}\,,
\end{aligned}\end{equation}
and $\thi_n=(1+s_n)/2$. Solving Eqs.~(\ref{ctymc}), we find that the reflection coefficients indeed have the form $r^\s_{n'n}=c^\s_{n}\de_{n'n}+d^\s_{n}\de_{n'\bar n}$, where
\begin{equation}
\label{AFMrnm}
c_{n}^\s=\frac{A^-_{n\s}A^+_{\bar n\s}-B^-_{n\s}B^+_{\bar n\s}}{A^+_{n\s}A^+_{\bar n\s}-B^+_{n\s}B^+_{\bar n\s}}\,,\,\,\,d^\s_n=-\frac{A^-_{\bar n\s}B^+_{\bar n\s}-B^-_{\bar n\s}A^+_{\bar n\s}}{A^+_{n\s}A^+_{\bar n\s}-B^+_{n\s}B^+_{\bar n\s}}\,,
\end{equation}
$A^\pm_{n\s}=\al^+_{n\s}\pm\al^-_{n\s}$, and $B^\pm_{n\s}=\be^+_{n\s}\pm\be^-_{n\s}$. One can verify that $c^+_n=c^-_n$ while $d^+_n=-d^-_n$. 

\subsection{Disordered interfacial exchange coupling}

This result can be generalized to the case when the exchange field felt by the electrons is disordered along the interface, i.e., $\eta_0\rightarrow\eta(i_y)\equiv\eta_0+\de\eta(i_y)$. Here, we consider a Gaussian-distributed disorder with zero mean and variance $V_\eta$, which is short-range correlated in the $y$ direction: $\langle\de\eta(i_y)\rangle=0$ and $\langle\de\eta(i_y)\de\eta(j_y)\rangle=V_\eta\de_{i_yj_y}$. The Hamiltonian on the magnetic side is now written in the $(n,n')$ basis as $H^\s_{nn'}=(-\hbar^2\pd_x^2/2m+U_0)\de_{nn'}+h^\s_{nn'}$, where $h^\s_{nn'}=E_n\de_{nn'}+\eta^\s_{nn'}$ and $\eta^\s_{nn'}=\sum_{i_y}\f_n(i_y)\s\eta(i_y)\f_{n'}(i_y)$. Diagonalizing $h^\s_{nn'}$ with an unitary matrix $W_\s$, i.e. $[W^\dag_\s h^\s W_\s]_{nn'}=d^\s_{nn'}=\la^\s_n\de_{nn'}$, the eigenenergies and eigenmomenta become $\ve^\s_{kn}=\hbar^2k^2/2m+U_0+\la^\s_n$ and $k_{n\s}^>=\sqrt{2m(E-U_0-\la^\s_n)}/\hbar$, respectively. The scattering state in the magnet can then be written as
\beq
\y^{>}_{nE\s}(x,i_y)=\sum_{n'}\frac{t^\s_{n'n}}{\sqrt{2\p\hbar v^>_{n'\s}}}\z_{n'\s}(i_y)e^{ik^>_{n'\s}x}\,,
\eeq
where $\z_{n\s}(i_y)=\sum_m (W^{\dag}_\s)_{nm}\f_m(i_y)$. We define two new $N\times N$ matrices, $X^\s_{nn'}=\sqrt{k^<_n/k^>_{n'\s}}(W^*_\s)_{nn'}$ and $Y^\s_{nn'}=\sqrt{k^>_{n'\s}/k^<_{n}}(W^*_\s)_{nn'}$. Imposing the continuity of wave function and its derivative at $x=0$ gives two matrix equations
\begin{equation}
{\boldsymbol 1}_N+\br^\s=\bX^\s\bt^\s\,,\,\,\,{\boldsymbol 1}_N-\br^\s=\bY^\s\bt^\s\,,
\end{equation}
where $\br$ and $\bt$ are the $N\times N$ reflection and transmission matrices, respectively, and ${\boldsymbol 1}_N$ is the $N\times N$ identity matrix. Solving for $\br^\s$, we obtain
\beq
\br^\s=-\left[{\boldsymbol 1}_N+\bX^\s(\bY^\s)^{-1}\right]^{-1}\left[{\boldsymbol 1}_N-\bX^\s(\bY^\s)^{-1}\right]\,.
\eeq
The spin-mixing conductance is then obtained from Eq.~(\ref{gudformula}) by summing over non-evanescent modes in the 2DEG.

\begin{figure}[t]
\centering
\includegraphics*[width=0.9\linewidth,clip=]{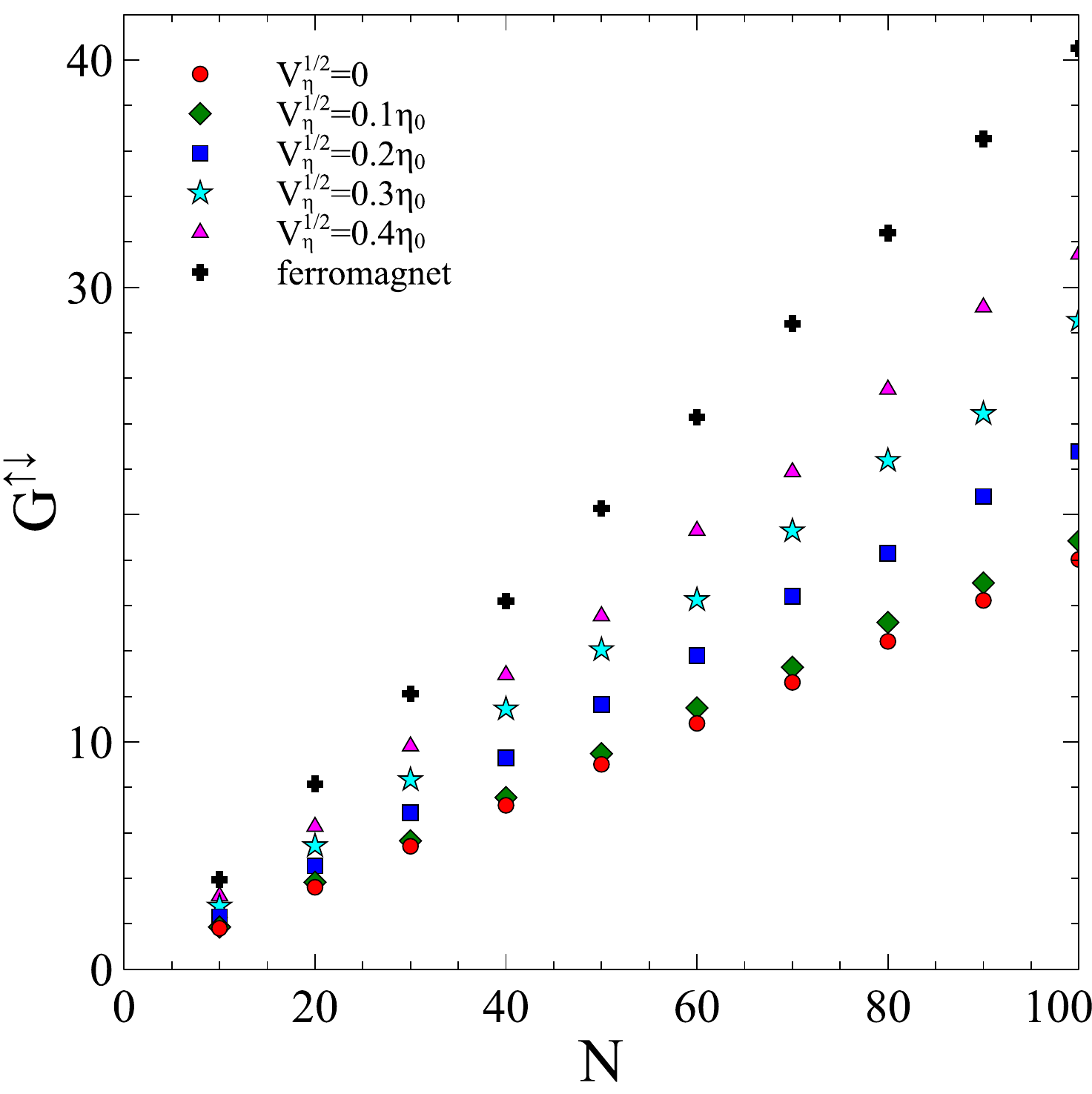}
\caption{Area-integrated spin-mixing conductance, $G^{\up\down}$, as a function of the system size, $N$. We have used $U_0=2.6W$, $\eta_0=0.9W$, $E=1.1W$. The red circles represent results for a perfectly uniform coupling $\eta_0$. For a disordered coupling $\eta(i_y)$ (see text for details), the spin-mixing conductance increases monotonically as the variance is increased. Here, we have used normalized standard deviations of $V^{1/2}_\eta/\eta_0=0.1, 0.2, 0.3, 0.4$. Disorder average was performed over 50 samples. The ``plus" points represent results for the ferromagnet$|$metal interface, where we have applied the formalism of this section with a (nonstaggered) uniform exchange field $\eta_0$.}
\label{mixcondplot}
\end{figure}

In Fig.~\ref{mixcondplot}, we have plotted the spin-mixing conductance for the uniform interfacial exchange coupling, $V_\eta=0$, as well as for finite disorder in the interfacial coupling, with standard deviations $V^{1/2}_\eta=0.1\eta_0$, $0.2\eta_0$, $0.3\eta_0$, and $0.4\eta_0$. Disorder averaging was performed over 50 samples. We see from Fig.~\ref{mixcondplot} that for both uniform and disordered coupling, the spin-mixing conductance, scales linearly as a function of the number of lattice sites $N$ (i.e., the interface area). As shown in the plot, the largest spin-mixing conductance (represented by ``plus" points) was obtained for a ferromagnet$|$metal interface, where we have applied the formalism above with $\eta_0(i_y)\equiv0.9W$. It is seen from the figure that as disorder variance $V_\eta$ is increased, $G^{\up\down}$ increases monotonically, and can reach values that are similar to the spin-mixing conductance of the ferromagnet$|$metal interface. This shows that despite full magnetization compensation in each of the antiferromagnetic unit cell, transfer of spins from a metal into an AF can be nearly as efficient as into a ferromagnet, especially in the presence of disorder. Recently, an experimental investigation showing large spin-mixing conductance at AF$|$metal interfaces, consistent with our expectation, was reported in Ref.~\onlinecite{merodioAPL14}, based on ferromagnetic resonance and spin pumping. Further experimental studies of the spin-mixing conductance for an AF$|$metal interface are desired.

\section{Summary and Discussion}
\label{conc}

In this work, we have theoretically discussed how spin-superfluid transport can be realized in an antiferromagnetic insulator. The phenomenon can be detected through a nonlocal-conductance measurement similar to the one proposed in Ref.~\onlinecite{takeiCM13}. Charge current $J^c_l$ along the $y$ axis in the left reservoir establishes, via the spin Hall effect, a spin accumulation at its interface to the AF, leading to the injection of spin into the AF. The spin density then propagates collectively through the AF and is pumped into the right reservoir, which can be detected by measuring the transverse charge current $J^c_r$ (or voltage, in an open circuit) generated in the negative $y$ direction through the inverse spin Hall effect (see Fig.~\ref{fig:setup}). The (negative) drag coefficient is then obtained as the ratio $\mathcal{D}\equiv J^c_r/J^c_l$. Using Onsager reciprocity and spin continuity at the interfaces, it was argued in the case of easy-plane ferromagnets that this drag coefficient can be orders of magnitude larger than the similar effect predicted for the incoherent transport of thermal magnons.\cite{takeiCM13} The same enhancement in the drag coefficient is also expected here. For comparable mixing conductances, we, furthermore, expect a similar magnitude for $\mathcal{D}$ in both AF- and ferromagnet-based heterostructures (which in Ref.~\onlinecite{takeiCM13} was estimated to be of order $0.1$ for a Pt$\mid$YIG$\mid$Pt sandwich when $L\lesssim L_\alpha$, where YIG stands for the yttrium iron garnet ferrimagnet).

Two promising materials directly relevant for the proposal presented here are antiferromagnetic insulators RbMnF$_3$ and KNiF$_3$. These insulators have a simple cubic perovskite structure both with a lattice constant of $a\approx4~\AA$,\cite{jiangJPCS06} and are known to be very good realizations of a nearest-neighbor isotropic 3D Heisenberg AF.\cite{jonghAIP74} RbMnF$_3$ is a spin $S=5/2$ AF with a N\'eel temperature of $T_N\approx83$K. The large spin moment associated with the Mn$^{2+}$ ions arises from a half-filled $3d$ electronic shell, and the nearest-neighbor antiferromagnetic exchange originates from the superexchange mechanism through the intervening F$^-$ ions. The nearest-neighbor exchange constant, extracted from inelastic neutron scattering, is $J\approx0.29$~meV, with the next-nearest-neighbor interaction measured to be more than an order of magnitude smaller than this value.\cite{coldeaPRB98} Antiferromagnetic-resonance measurements show magnetic anisotropy field of less than 10$^{-5}$ of the exchange field.\cite{teaneyPRL62,*freiserPRL63} KNiF$_3$ is an $S=1$ AF with a N\'eel temperature of $T_N\approx275$K. This material is unique in that it retains its cubic crystal symmetry down to temperatures well below $T_N$, while the other compounds in its class, e.g., KMnF$_3$, KFeF$_3$, KCoF$_3$, and KCuF$_3$, loose their cubic crystal symmetry at low temperatures.\cite{okazakiJPSJ59} The exchange constant is $J\approx8$~meV, and the magnetic anisotropy field is of the order 10$^{-5}$ of the exchange field.\cite{yamaguchiPRB99}

As mentioned below Eqs.~(\ref{mainres}), the transmission of spin current through the antiferromagnet is essentially lossless for system sizes $L\ll L_\al\equiv\hbar g^{\up\down}/4\p\al's$, while the loss becomes appreciable and the spin current decays algebraically as $L_\alpha/L$ for $L\gg L_\al$. We now give a quantitative estimate for the crossover length $L_\al$ between these two regimes. A reasonable upper bound for the spin-mixing conductance is $\lesssim10^{19}$~m$^{-2}$, an experimental value recently reported for several YIG$|$normal-metal interfaces,\cite{heinrichPRL11,*kurebayashiNATM11,*burrowesAPL12,*hahnPRB13} which shows that it is close to the ideal Sharvin limit.\cite{tserkovPRL02sp} Let us use a more conservative $g^{\up\down}\sim10^{18}$~m$^{-2}$ for our AF$|$metal interface, in light of our results in Sec.~\ref{afsmc}. An estimate for the Gilbert damping parameter $\al'$ is made based on a series of antiferromagnetic resonance (AFMR) experiments conducted on a body-centered tetragonal antiferromagnet, MnF$_2$. An early zero-field AFMR study on single-crystal slabs of MnF$_2$ has measured residual low-temperature ($T\approx 25$~K) normalized linewidth $\De\w/\w_{\rm res}\approx 6\times 10^{-3}$ with resonance frequency $\w_{\rm res}/2\p\approx 250$~GHz.\cite{johnsonPR59} This translates into a Gilbert damping parameter $\al'\sim (\De\w/\w_{\rm res})\sqrt{H_a/H_e}\sim7\times10^{-4}$, where the anisotropy and exchange fields for MnF$_2$ are given by $H_a\sim8$~kOe and $H_e\sim 500$~kOe, respectively.~\cite{johnsonPR59,kotthausPRL72} From a later high-field AFMR experiment using flat disk samples,\cite{kotthausPRL72} where the resonance frequencies are driven down into the mm-wave region, an improved normalized linewidth $\De\w/\w_{\rm res}\approx6\times10^{-4}$ was reported with $\w_{\rm res}/2\p\approx23$~GHz at $T\approx 4$~K,~ translating into $\al'\sim 8\times10^{-5}$. Ref.~\onlinecite{kotthausPRL72} argues that the narrower linewidth observed for their uniform mode compared to the 250-GHz experiment is due to a weaker scattering into magnetostatic modes that are degenerate with the uniform mode at lower frequencies. We also note here that the Gilbert damping parameter for YIG has been experimentally reported to be\cite{heinrichPRL11} $\al'\lesssim10^{-4}$, a value with potential relevance for us because of the ferrimagnetic nature of YIG with partially compensated magnetic moments inside its unit cell. Based on these numbers, we take, somewhat conservatively, $\al'\sim10^{-4}$ for our estimate of $L_\al$. Using $S=5/2$ and lattice constant $a\approx4~\AA$ corresponding to RbMnF$_3$, the crossover length reads $L_\al\sim20$~nm. For KNiF$_3$, with $S=1$ and $a\approx 4~\AA$, we obtain $L_\al\sim50$~nm. A lower $\alpha'$ (which could be expected for relevant cubic materials that have much weaker anisotropy) and/or larger $g^{\up\down}$ could possibly push $L_\alpha$ into the $\mu$m range.

In-plane magnetic anisotropy, which breaks the U(1) symmetry of the magnetic order, can, furthermore, lead to pinning of the N\'eel order parameter and thus quench the global precession essential for superfluid transport. The anisotropy therefore defines a critical spin-current density that needs to be injected at the interface in order that the N\'eel vector overcomes the pinning. An estimate for this critical spin current can be made by supplementing Lagrangian density (\ref{LAF}) with an in-plane easy-axis anisotropy term of the form $\ka n_x^2/2$, where $\ka=\gamma sH_a$ parametrizes its strength and $n_x$ is the $x$ component of the unit N\'eel vector. For sufficiently small $L$, the antiferromagnet can be treated as a monodomain. Then the (volume-integrated) N\'eel vector is subjected to a restoring torque $\sim\ka\mathscr{V}_{\rm AF}$ as it rotates away from the easy axis in the plane. This pinning torque needs to be overcome by the spin torque from the spin Hall metal, $\mathscr{A}J^s$. The critical spin-current density in this regime is then given by
\beq
\label{Jsc}
J^s_c\sim\ka L\,.
\eeq 
This critical spin current is proportional to length $L$, because the spin torque is generated interfacially whereas the pinning is taken to be due to a bulk anisotropy. However, as $L$ increases above a certain crossover length $L_c$, the critical spin current can be expected to saturate as a function of $L$. This crossover length can be self-consistently shown to obey $L_c=2\p A(J^s_c)^{-1}|_{L=L_c}$ (corresponding to the helical pitch of an isotropic AF subjected to spin current $J_c^s$), and is thus given by $L_c=\sqrt{2\p A/\ka}$. For $L\gg L_c$, the critical spin-current density thus reads
\beq
\label{Jsc>}
\tilde{J}^s_c\sim\kappa L_c=\sqrt{2\p A\ka}\,.
\eeq
From a complementary perspective, this threshold corresponds to the spin supercurrent carried by a static magnetic texture maintained by appropriate boundary conditions, at which the domain walls that separate regions with opposite N\'eel orientations along the easy axis begin to coalesce.\cite{konigPRL01} 

We now make some quantitative estimates for the simple cubic nearest-neighbor antiferromagnets, RbMnF$_3$ and KNiF$_3$. In this case, the stiffness parameter entering Eq.~(\ref{LAF}) is given by $A=JS^2/a$. For RbMnF$_3$, we get $A\approx 7\times10^{-13}$~J$\cdot$m$^{-1}$, and, using $H_a\approx4.5$~Oe,\cite{jonghAIP74} the crossover length scale reads $L_c\approx100~\mbox{nm}$. The critical spin-current density (\ref{Jsc>}) is estimated to be $\tilde{J}^s_c\sim10^{-5}$~J$\cdot$m$^{-2}$. This is converted into the applied electric-current density $J^c$ according to $J^s=(\hbar/2e)\theta_{\rm SH}J^c$, where $\theta_{\rm SH}$ is the effective electron spin Hall angle at the metal$\mid$AF interface. Using $\theta_{\rm SH}\sim0.1$ appropriate for a platinum contact,\cite{liuPRL11} the necessary current density then becomes $J^c\sim5\times10^{11}~\mbox{A}\cdot\mbox{m}^{-2}$. For KNiF$_3$, $A\approx3\times10^{-12}$~J$\cdot$m$^{-1}$, and, using $H_a\approx 80$~Oe,\cite{yamaguchiPRB99} the crossover length in this case is given by $L_c\approx90$~nm. The critical spin-current density reads $\tilde{J}^s_c\sim2\times10^{-4}$~J$\cdot$m$^{-2}$, increasing the electric current to $J^c\sim7\times10^{12}$~A$\cdot$m$^{-2}$ for the platinum contact.

The focus of the work has been the low-temperature regime, where spin is carried through the AF mostly by the condensate. In contrast to easy-plane ferromagnets,\cite{takeiCM13} however, spin transport in AFs poses a much richer problem at higher temperatures, where the thermal cloud can contribute appreciably to the transport. In the case of the ferromagnet with an easy $xy$ plane, magnons are not capable of carrying spin current polarized along the $z$ axis in the bulk of the magnet. Injection of spin into the thermal cloud at the ferromagnet$|$metal interface quickly converts into the condensate over a healing length $\sim\sqrt{A/K}$, where $A$ and $K$ are respectively ferromagnetic stiffness and easy-plane anisotropy, and the spin is carried by the condensate in the bulk. A theory of finite-temperature spin transport in AFs, on the other hand, has two-fluid character involving the condensate and the thermal-cloud contributions. In the long-wavelength limit within the collision-dominated regime, for example, insight into spin transport through AFs should be obtainable from the two-fluid hydrodynamic theory developed for superfluid $^4$He by Landau and Khalatnikov.\cite{khalatnikovBOOK65} We therefore anticipate that the understanding of finite-temperature spin transport through AFs will further deepen the connection between spin superfluidity and conventional mass superfluidity.

\acknowledgments

ST and YT would like to thank Mircea Trif for valuable discussions and acknowledge support in part by FAME (an SRC STARnet center sponsored by MARCO and DARPA), the NSF under Grant No. DMR-0840965, and the Kavli Institute for Theoretical Physics through Grant No. NSF PHY11-25915. BIH and AY acknowledge support in part by the STC Center for Integrated Quantum Materials under NSF Grant No. DMR-1231319.

\end{document}